\documentstyle[12pt,a4]{article}

\begin{document}

\title{\sc MONOPOLE CONDENSATES IN SEIBERG-WITTEN THEORY }
\author{Cihan Sa\c{c}l\i o\~{g}lu$^{1,2}$\\
\date{$^{1}$Physics Department, Bo\~{g}azi\c{c}i University \\
80815 Bebek--\.{I}stanbul, Turkey\\
and \\
$^{2}$Feza G\"{u}rsey Institute \\   TUBITAK--Bo\~{g}azi\c{c}i University\\ 
81220 \c{C}engelk\"{o}y, {I}stanbul-- Turkey}}
\maketitle
\vspace*{1cm}
\begin{abstract}
A product of two Riemann surfaces of genuses $p_{1}$ and 
$p_{2}$ solves the Seiberg-Witten monopole equations for
a constant Weyl spinor that represents a monopole condensate.  
Self-dual electromagnetic fields require $p_{1} = p_{2}= p$ and provide a 
solution of the euclidean Einstein-Maxwell-Dirac equations with
$p-1$ magnetic vortices in one surface and the same number of 
electric vortices in the other.  The monopole condensate plays 
the role of cosmological constant.  The virtual dimension of the 
moduli space is zero, showing that for given $p_{1}$ and $p_{2}$, 
the solutions are unique. 
\end{abstract}
\vspace*{3 cm}
\pagebreak
\baselineskip=30pt

In this note we present solutions of the Seiberg-Witten monopole equations (SWME)\cite{Wit1}
\begin{equation}\label{1}
\not{\! \! D}_{A} \psi \equiv \gamma^{a} E^{\mu}_{a}
(\partial_{\mu}+iA_{\mu}+\frac{1}{8}
\omega^{bc}_{\mu}[\gamma_{b},\gamma_{c]})\psi=0,~~.
\end{equation}
\begin{equation}\label{2}
F^{+}_{\mu\nu}\equiv\frac{1}{2}(F_{\mu\nu} + 
\frac{1}{2}\epsilon_{\mu \nu \alpha \beta}F^{\mu\nu}) =-\frac{i}{4} \psi^{\dag}[\gamma_{\mu},\gamma_{\nu}]\psi~~~
\end{equation}
for which the 4-manifold ${\cal M}_{4}$ has the form 
$\Sigma_{p_{1}} \times \Sigma_{p_{2}}$ with 
$p_{1}+p_{2}\geq 2$, excluding $p_{1}=p_{2}=1$.  
The Weyl spinor $\psi$, which represents  massless monopoles in
the SWME, consists here of a single constant component
$\psi_{1} (\psi_{2})$, giving rise to a monopole (antimonopole)
condensate.  Physically, $p_{1} -1~(p_{2}-1)$ is the number 
of magnetic (electric) vortices in $\Sigma_{p_{1}}~( \Sigma_{p_{2}})$.  
Remarkably, the most symmetric case with self-dual electromagnetic 
fields is also of direct physical interest: it is a solution of the coupled 
Einstein-Maxwell-Dirac equations, with the condensate now serving as the cosmological constant.  The field equations are

\begin{equation}\label{3}
{\cal R}_{\mu\nu}-\frac{1}{2}g_{\mu\nu}{\cal R}=
\kappa(T_{\mu\nu}(\mbox{\small e.m.})+
T_{\mu\nu}(\mbox{\small Dirac}))+\Lambda g_{\mu\nu},
\end{equation}
\begin{equation}\label{4}
F^{\mu\nu}_{;\mu}=0,
\end{equation}
\begin{equation}\label{5}
\tilde{F}^{\mu\nu}_{;\mu}=0~~,
\end{equation}
\noindent together with the Dirac equation (\ref{1}).  
In (\ref{1})-(\ref{5}),
 $ \omega^{bc}_{\mu}$ and $A_{\mu}$
are  the spin and $U(1)$ connections.  The  $\gamma^{a}$
are the flat-space $\gamma-$matrices; thus $\gamma^{\mu}e^{a}_{\mu}=\gamma^{a}$ 
with $E^{a}_{\mu}e_{a}^{\nu}=\delta_{\mu}^{\nu},~ e^{a}_{\mu} \delta_{ab}e^{b}_{\nu}=
g_{\mu\nu}$, etc.  
The Weyl spinor $\psi$ is of the 
form $\psi^{T}=(\psi_{1}, \psi_{2}, 0, 0)$  and the matrices

\noindent $\gamma_{a}=\tau_{1}\bigotimes\sigma_{a}(a=1, 2, 3), 
\gamma_{4}=\tau_{1}\bigotimes\bf{1}$ are block off-diagonal, while 
$\gamma_{5}=\tau_{3}\bigotimes\bf{1}$. Since the metric is Euclidean, $\psi^{\dag}$ replaces $\bar{\psi}$ 
in all spinor bilinears.  These lead to the automatic vanishing
of spinor currents $\psi^{\dag}\gamma_{\mu}\psi $ or  $ \psi^{\dag}\gamma_{\mu}\gamma_{5}\psi$ which would otherwise 
have appeared on the RHS of (\ref{4}) and (\ref{5}).

The cosmological constant is needed 
since non-singular and square-integrable solutions of the SWME with
${\cal R}(x)\geq 0$  are forbidden by the Weitzenbock formula and Witten's vanishing theorems, which are essentially its integrated version (for a review of the SWME, see \cite{Akb}).
As $ T^{\mu}_{\mu}(\mbox{\small e.m.})= T^{\mu}_{\mu}(\mbox{\small Dirac})=0$,
we are clearly limited to manifolds ${\cal M}_{4}$ where $4\Lambda=-{\cal R}$
is a positive constant, which will turn out to be provided by the monopole condensate.

It is sensible to begin with a euclideanized Bertotti-Robinson \cite{Bert}, \cite{Rob} type Ansatz,  
as that solution also involves covariantly constant electromagnetic fields and admits a cosmological  constant \cite{Bert}. 
Thus, as in \cite{Bert}, we may try 
${\cal M}_{4}={\cal M}^{(1)}_{2}\times{\cal M}^{(2)}_{2}$.  
One can then choose  the conformally-flat basis one-forms
\begin{equation}\label{6}
e^{i}=e^{\mu}dx^{i},  \mu=\mu(x^{1},x^{2}), i=1, 2;~~ e^{j}=e^{\nu}dx^{j},  
\nu=\nu(x^{3},x^{4}), j=3,4 
\end{equation}
\noindent  and
\begin{equation}\label{7}
A_{\mu}=(A_{1}(x^{1},x^{2}), A_{2}(x^{1},x^{2}),A_{3}(x^{3},x^{4}),A_{4}(x^{3},x^{4}))
\end{equation}
with no initial restriction on $\psi$.  Putting the Ansatz in the 
SWME , however, one finds that
one of $\psi_{1}$ or $\psi_{2}$ must be zero, while the other can at most be
a non-vanishing constant.  The cases $\psi_{1}\neq0$ and $\psi_{2}\neq0$
can be regarded as condensates of massless monopoles or 
antimonopoles, respectively.  This is reminiscent of, and possibly dual to the gluino condensate 
considered by Witten \cite{Wit2} in the original topological 
twisted $N=2$ supersymmetric Yang-Mills
theory.  

Let us start with $\psi_{1}\neq0$.  This yields ${\cal R}=2{\cal R}^{12}_{12} 
+2{\cal R}^{34}_{34}= -2 |\psi_{1}|^{2}  $.  Introducing a new constant, we put
 ${\cal R}^{12}_{12} = - |\phi|^{2}$; thus ${\cal R}^{34}_{34}=  
 -(|\psi_{1}|^{2}- |\phi|^{2})$.  We now have three possibilities: 
(i) Both $ {\cal M}_{2} $'s have constant negative curvatures ($|\psi_{1}|>|\phi|$);
(ii) one $ {\cal M}_{2}$ is flat while the 
other has constant negative curvature ($|\phi|=0$ or $|\psi_{1}|=|\phi|$);  (iii) $|\psi_{1}|<|\phi|$, 
hence ${\cal M}^{(2)}_{2}=S^{2} $ and ${\cal M}^{(1)}_{2} $ has constant 
negative curvature.  In all the cases, use of the Cartan structure equations, 
the constancy of $|\psi_{1}|$ and the SWME lead to

\begin{equation}\label{8}
(A_{1}, A_{2})=-\frac{1}{2}(\omega^{12}_{1},\omega^{12}_{2});~~ (A_{3}, A_{4})=-\frac{1}{2}(\omega^{34}_{3},\omega^{34}_{4})
\end{equation}
\noindent and therefore 

\begin{equation}\label{9}
F=dA=-d\frac{1}{2}(\omega^{1}_{2}+\omega^{3}_{4})=-\frac{1}{2}(R^{1}_{2}+R^{3}_{4}),
\end{equation}

\noindent since the curvature two-form 
${\cal R}^{a}_{b}={\cal R}^{a}_{bcd}e^{c} \wedge e^{d}$ contains no $\omega^{a}_{c} \wedge \omega^{c}_{b}$ terms in our Ansatz.

It is worth pausing briefly to consider some implications of (\ref{8}) and (\ref{9}).  
First of all, (\ref{8}) and the form of $\psi$ show that the original version 
of Hermann Weyl's "Eichinvarianz" \cite {Weyl} actually holds 
in (\ref{1}): A gauge transformation 
$A_{\mu} \rightarrow A_{\mu} +  \partial_{\mu}\alpha(x^{1}, x^{2}) 
+ \partial_{\mu}\beta (x^{3}, x^{4}) $ can be compensated for by local changes 
of the scale factors $\mu$ and $\nu$ !  Secondly, this gauge transformation
reveals that each of the two 2-manifolds effectively has its own $U(1)$ fiber.  
Thirdly, $ (D_{A})_{\mu}\psi=0$ 
by (\ref{8}); hence $T_{\mu\nu}(\mbox{\small Dirac})=0$ in (\ref{3}).  Then, 
contracting (\ref{3}), one finds $-{\cal R} =2 |\psi_{1}|^{2}=4\Lambda$; hence, as mentioned before, 
{\em the cosmological 
constant is given by the monopole condensate}.  Incidentally, this  
represents a counterexample to the folklore that solutions of the 
coupled Einstein-Maxwell-Dirac equations need not be sought
because the Dirac field is supposed to become negligible in the classical limit
where the Einstein-Maxwell equations apply.

We now seek solutions of the SWME for  the cases (i)-(iii).  It will be seen later that 
the most highly symmetric and self-dual special case of (i) also solves 
(\ref{3}-\ref{5}).  We define the pair of complex coordinates 
$z^{1}\equiv x+iy\equiv\sqrt{2}|\phi|(x^{1}+ix^{2})$,
$z^{2}\equiv s+it\equiv\sqrt{2(|\psi_{1}|^{2}-|\phi|^{2})}(x^{3}+ix^{4})$.  The SWME
then result in
\begin{equation}\label{10}
4\partial_{2} \partial_{\overline{2}}\nu = 
e^{2\nu}~~~(|\psi_{1}|>|\phi|),
\end{equation}
\begin{equation}\label{11}
4\partial_{2} \partial_{\overline{2}}\nu = 
0~~~(|\psi_{1}|=|\phi|),
\end{equation}
\begin{equation}\label{12}
4\partial_{2} \partial_{\overline{2}}\nu = 
-e^{2\nu}~~~(|\psi_{1}|<|\phi|)
\end{equation}
\noindent for (i), (ii) and (iii), respectively.  In addition, $\mu$ satisfies
\begin{equation}\label{13}
4\partial_{1} \partial_{\overline{1}}\mu = 
e^{2\mu} 
\end{equation}
in all three cases.  Now it is known that the Liouville equation
\begin{equation}\label{14}
4\partial_{z} \partial_{\overline{z}}\Lambda = 
\pm e^{2\Lambda} 
\end{equation}
has the general solution \cite{Liou}
\begin{equation}\label{15}
\Lambda=\frac{1}{2} \ln \frac{|dg/dz|^{2}}{(1\mp g\overline{g})^2},
\end{equation}
where $g(z)$ is an arbitrary analytic function.  Using (\ref{8}), (\ref{9}) 
and (\ref{15}), the solutions of (1) and (2) now may be summarized as (we repeat
(\ref{8}) and (\ref{9}) in order to display all the results together)
\begin{equation}\label{16}
\omega^{12}= -i \{ \frac{1}{2}d\ln (\frac{d\overline{g}_{1}}{d\overline{z}_{1}}
\frac{dz_{1}}{dg_{1}}) + 
\frac{(g_{1}d\overline{g}_{1} - \overline{g}_{1}dg_{1})}{(1-g_{1}\overline{g}_{1})}\},
\end{equation}

\begin{equation}\label{17}
\omega^{34}= -i \{ \frac{1}{2}d\ln (\frac{d\overline{g}_{2}}{d\overline{z}_{2}}
\frac {dz_{2}}{dg_{2}}) + 
\frac {(g_{2}d\overline{g}_{2}- \overline{g}_{2}dg_{2})}{(1 \mp g_{2}\overline{g}_{2})} \},
\end{equation}

\begin{equation}\label{18}
A=-\frac{1}{2}(\omega^{1}_{2}+\omega^{3}_{4}),
\end{equation}

\begin{equation}\label{19}
R^{1}_{\: 2}  =  -2i\frac{dg_{1} \wedge d \overline{g}_{1}}{(1-g_{1}\overline{g}_{1})^2};
~R^{3}_{\: 4}=  -2i\frac{dg_{2} \wedge d \overline{g}_{2}}{(1 \mp g_{2}\overline{g}_{2})^2},
\end{equation}

\begin{equation}\label{20}
F=-\frac{1}{2}(R^{1}_{2}+R^{3}_{4}),
\end{equation}
\noindent where the upper and lower signs in (\ref{17}) and (\ref{19}) 
correspond to cases (i) and (iii), i.e., negative and positive constant 
curvature for ${\cal M}^{(2)}_{2}$.  For case (ii), we may solve (\ref{11})
by taking $\nu=\zeta (z_{2}) + \overline{\zeta}(\overline{z}_{2})$, which 
gives ${\cal M}^{(2)}_{2}$ the flat metric 

\begin{equation}\label{21}
ds^{2}=\exp {(\zeta (z_{2}) + \overline{\zeta}(\overline{z}_{2}))} 
dz_{2}d\overline{z}_{2} \equiv dw_{2}d\overline{w}_{2};
\end{equation}
thus we can use (\ref{16})-(\ref{19}) with $\omega^{3}_{4}=R^{3}_{\: 4}=0. $

So far we have only mentioned the curvatures of the two manifolds; 
their global topological properties depend on the choices for $g_{1}(z_{1}),
g_{2}(z_{2})$ and $w(z_{2})$.  For example, taking $w(z_{2})$ to be an
inverse elliptic function makes ${\cal M}^{(2)}_{2}$ the genus $p=1$ flat
torus.  Similarly, but less trivially, we can tesellate the constant
negative curvature hyperboloid $|g_{1}(z_{1})|\leq 1$ by 
$4p_{1}$-gons 
with geodesic edges. Pairwise identifying the latter
in the usual way \cite{Novikov} turns ${\cal M}^{(1)}_{2}$  into a genus 
$p_{1}$ Riemann surface $\Sigma_{p_{1}}$.  To do this explicitly, 
one first goes over to the Poincar\'{e} metric for the upper half-plane 
$C_{+}$

\begin{equation}\label{22}
ds^{2}= \frac {df_{1}d\overline{f}_{1}}{(Im f_{1})^{2}}
\end{equation}

\noindent where  $g_{1}(z_{1})=(f_{1}-i)/(f_{1}+i) $, and chooses $f_{1}$
as the Fuchsian function \cite{Nehari}, \cite {Ford} used in uniformizing an algebraic 
function whose Riemann surface has genus $p_{1}$.  Obviously,
in case (i) a similar choice can be made for $g_{2}(z_{2})$, giving
 ${\cal M}^{(2)}_{2}=\Sigma_{p_{2}}$.  In case (iii), we have 
 ${\cal M}^{(2)}_{2}=\Sigma_{0} \equiv S^{2}$ and the only one-to-one 
and onto mappings of $S^{2}$ to itself consist of 
$g_{2}(z_{2})=(az_{2}+b)/(cz_{2}+d)$.  

To summarize, we have found
solutions of the SWME of the form 
${\cal M}_{4}=\Sigma_{p_{1}} \times \Sigma_{p_{2}}$ where 
$p_{1}+p_{2}\geq 2$, excluding $p_{1}=p_{2}=1$.  Using (\ref{18}),
the first Chern classes of the two manifolds are seen to be
$p_{1}-1$ and $p_{2}-1$; these  may be regarded as the number
of magnetic vortices on  ${\cal M}^{(1)}_{2}$ and 
electric vortices on ${\cal M}^{(2)}_{2}$, respectively.  In the 
"antimonopole condensate" case with $\psi_{1}=0,\psi_{2}\neq  0$,
the metric remains the same but $A_{\mu}$ and the first 
Chern classes change sign; this can be thought of as 
changing the sense of the vortices.

Are these solutions unique?  One may first think of generating new
solutions in cases (i) and (iii) via the metric-preserving 
transformations

\begin{equation}\label{23}
\tilde{g_{1}} = \frac{\alpha_{1} g_{1}+\beta_{1}}
{\overline{\beta_{1}}g_{1}+\overline{\alpha_{1}}},~\tilde{g_{2}} =
\frac{\alpha_{2} g_{1}+\beta_{2}}
{\pm \overline{\beta_{2}}g_{2}+\overline{\alpha_{2}}}, 
~~{\rm with}~~|\alpha_{i}|^2 \mp  |\beta_{i}|^2 = 1~~,
\end{equation}
\noindent where the upper (lower) sign again refers to (i) ((iii)).  
However, all the fields are seen to be invariant under such 
$SU(1,1)\times SU(1,1)$ $(SU(1,1)\times SU(2))$ transformations; in particular,
integer subgroups of the above only shuffle the $4-p_{i}$-gons, keeping the tesellation  
fixed.  To investigate uniqueness more generally, one needs to compute 
the virtual dimension $W$ of the moduli space of the solutions via

\begin{equation}\label{24}
W = - (2\chi + 3\sigma)/4 + c^{2}_{1},
\end{equation}
\noindent where the signature
\begin{equation}\label{25}
\sigma ({\cal M}_{4})= - \frac{1}{24 \pi^{2}}
\int_{{\cal M}_{4}} R^{a}_{b}\wedge R^{b}_{a}
\end{equation}
obviously vanishes for our solutions  in which only $R^{1}_{212}$ and 
$R^{3}_{434}$ are non-zero.  The Euler characteristic follows from 
the Kunneth formula
\begin{equation}\label{26}
\chi ({\cal M}_{4})=
\chi ({\cal M}^{(1)}_{2}) \chi ({\cal M}^{(2)}_{4})=(2-2p_{1})(2-2p_{2}),
\end{equation}
\noindent while, using (\ref{20}), one sees that

\begin{equation}\label{27}
 c^{2}_{1}={1\over (2\pi )^{2}} \int F^{2}= \chi ({\cal M}_{4})/2 .
\end{equation} 

\noindent Hence $W=0$, proving that our SWME solutions are unique up to 
gauge and conformal transformations. 

We return to the question of which of the above solutions 
also satisfy the remaining "physical" equations, of which (\ref{3}) has now 
been reduced to

\begin{equation}\label{28}
{\cal R}_{\mu\nu}-\frac{1}{2} g_{\mu\nu} {\cal R}=
\kappa(T_{\mu\nu}(\mbox{\small e.m.}))+
\frac{1}{2}|\psi_{1}|^{2} g_{\mu\nu}.
\end{equation}
\noindent Direct computation shows that only
case (i) is compatible with (\ref{28}) for the special value
$2|\phi|=|\psi_{1}|$, which makes the curvatures of the two manifolds 
equal. By (\ref{8}) and (\ref{9}), this means the $U(1)$ field strengths are 
self-dual. Hence (\ref{4}) and (\ref{5}) are both solved and $T_{\mu\nu}(\mbox{\small e.m.})=0$.  The Riemannian curvature
is also self-dual in the sense that the the equal sharing of the monopole 
condensate between the two manifolds results in their having the same curvature 
and the same genus; the only difference is that the vortices are "magnetic" in one
and "electric" in the other.  It would be interesting to investigate whether
(\ref{3}-\ref{5}) always select the most symmetric subset of the SWME
solutions.
   
\noindent {\bf Acknowledgements}

I am grateful to S. Akbulut for patient instruction 
in Seiberg-Witten theory, to P. Argyres for clearing 
a confusing point involving (\ref{27}), and to Y. Nutku 
for pointing out the similarities to the Bertotti-Robinson solution, 
carrying out the REDUCE check and editing.  
I also thank M. Ar\i k, T. Dereli and R. G\"{u}ven for helpful 
discussions and S. Nergiz for LATEX help.

\end{document}